\documentclass[letterpaper,12pt]{article}

\usepackage{array}
\usepackage{jheppub}
\usepackage[all,pdf]{xy}

\newcommand{\an}[1]{\langle{#1}\rangle}

\DeclareMathOperator{\Gr}{G}

\title{
Symbol Alphabets from Plabic Graphs II:\\
Rational Letters
}

\author[1]{J.~Mago,}
\author[1]{A.~Schreiber,}
\author[1,2]{M.~Spradlin,}
\author[1]{A.~Yelleshpur Srikant}
\author[1]{and A.~Volovich}

\affiliation[1]{Department of Physics,
Brown University,
Providence, RI 02912, USA}

\affiliation[2]{Brown Theoretical Physics Center,
Brown University,
Providence, RI 02912, USA}

\abstract{
Symbol alphabets of $n$-particle amplitudes in $\mathcal{N} = 4$ super-Yang-Mills
theory are known to contain certain cluster variables of $\Gr(4, n)$ as well as certain
algebraic functions of cluster variables.
The first paper arXiv:2007.00646 in this series focused on $n=8$ algebraic letters.
In this paper we show that
it is possible to obtain all rational symbol letters (in fact all cluster variables)
by solving matrix equations of the form
$C\, Z = 0$ if one allows $C$ to be an arbitrary
cluster parameterization of the top cell of $\Gr_+(n{-}4,n)$.
}

\begin{document}

\maketitle

\section{Introduction}

It was observed some time ago~\cite{Golden:2013xva} that many
symbol letters~\cite{Goncharov:2010jf}
of perturbative $n$-particle amplitudes
scattering amplitudes in planar maximally
supersymmetric Yang-Mills (SYM) theory
are cluster variables
of the $\Gr(4,n)$ cluster algebra~\cite{FZ1,FZ2,Scott,FWZ}.
Recently there has been renewed interest in the problem of
explaining, or possibly
even predicting,
the symbol alphabet for a given amplitude\footnote{The connection
between cluster algebras and symbol alphabets is of course
an interesting question also beyond SYM theory; see for
example~\cite{Chicherin:2020umh}.}.
Knowledge of symbol alphabets is important input for a bootstrap
program that is currently the state-of-the art for high-loop
calculations in SYM theory (see~\cite{Caron-Huot:2020bkp} for a review).
Some information on symbol alphabets can be gleaned~\cite{Dennen:2015bet,Dennen:2016mdk,Prlina:2017azl,Prlina:2017tvx,Prlina:2018ukf} by analyzing Landau equations, but
as noted in those references, this analysis cannot completely determine symbol letters.
More recent `phenomenological' approaches to this problem involving
tropical fans, polytopes, and plabic graphs associated to Grassmannians have been discussed
in~\cite{Arkani-Hamed:2019rds,Drummond:2019cxm,Henke:2019hve,Mago:2020kmp,Drummond:2020kqg,He:2020uhb}.
Motivation for the renewed interest comes in part
from recent calculations of new amplitudes~\cite{Zhang:2019vnm,He:2020vob}
that have provided fresh data on the sets of cluster variables
and algebraic functions of cluster variables that appear as symbol
letters of certain amplitudes.

In~\cite{Mago:2020kmp,He:2020uhb}
the connection between symbol letters and plabic graphs was explored
via matrix equations of the form $C \, Z=0$ that arise
naturally~\cite{ArkaniHamed:2009dn,Mason:2009qx,ArkaniHamed:2009vw,Drummond:2010uq}
in the Grassmannian formalism for amplitudes in SYM theory.
Here $C$ is a $k \times n$ matrix parameterizing a $4k$-dimensional
cell of the totally non-negative Grassmannian $\Gr_+(k,n)$~\cite{Postnikov,ArkaniHamed:2012nw}
and $Z$ is an $n \times 4$ matrix of momentum twistors~\cite{Hodges:2009hk}
specifying the kinematic data for $n$-particle scattering.
This paper should be read as a close sequel of~\cite{Mago:2020kmp}, to which
we refer the reader for many important details.
A highlight of that paper was the highly nontrivial appearance (also
found in~\cite{He:2020uhb}), for $k=2$,
of 18 independent \emph{algebraic} functions of $\Gr(4,8)$ cluster variables
that precisely match the 18
algebraic letters of the eight-particle symbol alphabet of~\cite{Zhang:2019vnm}.
However, these papers left open the question
of whether all \emph{rational} symbol letters
could be obtained by solving
$C \, Z = 0$.

In this paper we show that it is \emph{not} possible to obtain
all rational symbol letters from plabic graphs, but instead we show how
to obtain all rational letters by allowing non-plabic $C$-matrices.
If only plabic $C$-matrices are used, then 16 out of the
180 rational letters of the eight-particle
symbol alphabet cannot be obtained (see also~\cite{HL}).
Non-plabic $C$-matrices can be constructed by starting with a plabic graph
$C$-matrix and performing a ``non-square move'', i.e., by mutating on a node
of valence greater than four in the quiver that is dual to the graph\footnote{The part of
a cluster algebra that can be reached by only performing square moves
in the dual graph was called the \emph{accessible} algebra in~\cite{Paulos:2014dja}.}.
More specifically,
we show that
if one allows $C$ to be an arbitrary
$4(n{-}4)$-dimensional
cluster parameterization of $\Gr_+(n{-}4,n)$
(i.e., of the top cell)
then \emph{every} cluster variable of $\Gr(4,n)$ (in particular,
all rational symbol letters) can be obtained
by solving $C \, Z = 0$. Moreover one will \emph{never} encounter
non-cluster variables,
which is interesting in light of the observation made in~\cite{Mago:2020kmp,He:2020uhb}
that for some non-top cells, solving $C\,Z = 0$
yields quantities that are polynomial in Pl\"ucker coordinates
but are \emph{not} cluster variables.

The plan of the paper is as follows.
In Sec.~2 we explain the construction of non-plabic
cluster
parameterizations by way of an $n=7$ example and then
discuss how the rational letters of the eight-particle
symbol alphabet arise from solving $C \, Z = 0$.
In Sec.~3 we establish the claims of the previous
paragraph by observing that
solving $C \, Z = 0$ for top cells induces the obvious
$\Gr(n{-}m,n) \cong \Gr(m,n)$ cluster isomorphism.

\section{Non-Plabic Parameterizations}

\subsection{A Motivational Example}

\begin{figure}
\centering
\begin{tabular}
{
>{\centering\arraybackslash} m{0.3\textwidth}
>{\centering\arraybackslash} m{0.1\textwidth}
>{\centering\arraybackslash} m{0.3\textwidth}
}
\includegraphics[width=1.8in]{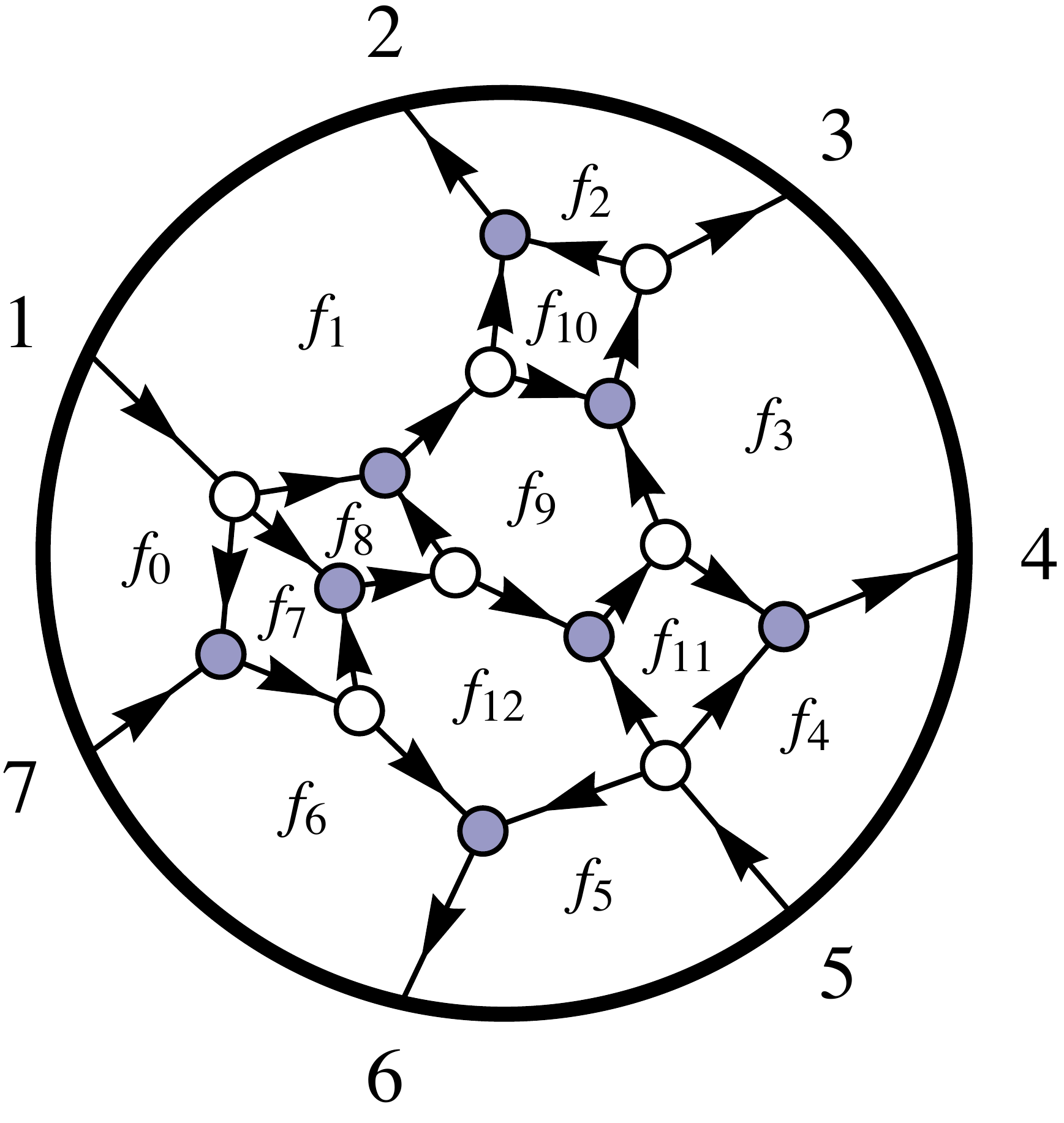}
&&
\begin{xy} 0;<60pt,0pt>:<0pt,60pt>::
(-0.6,0.6) *+{\framebox[3ex]{$\scriptstyle{f_1}$}} = "a",
(0.5,\halfrootthree) *+{\framebox[3ex]{$\scriptstyle{f_2}$}} = "b",
(1,-0.4) *+{\framebox[3ex]{$\scriptstyle{f_4}$}} = "c",
(1,0.4) *+{\framebox[3ex]{$\scriptstyle{f_3}$}} = "j",
(0.5,-\halfrootthree) *+{\framebox[3ex]{$\scriptstyle{f_5}$}} = "d",
(-0.5,-\halfrootthree) *+{\framebox[3ex]{$\scriptstyle{f_6}$}} = "e",
(-1,0) *+{\framebox[3ex]{$\scriptstyle{f_0}$}} = "f",
(0.25,0.4) *+{\scriptstyle{f_{10}}} = "g",
(0.0,0.0) *+{\scriptstyle{f_9}} = "h",
(-0.15,-0.5) *+{\scriptstyle{f_{12}}} = "i",
(-0.65,-0.4) *+{\scriptstyle{f_{7}}} = "k",
(-0.4,0.0) *+{\scriptstyle{f_{8}}} = "l",
(0.3,-0.3) *+{\scriptstyle{f_{11}}} = "m",
"e", {\ar"i"},
"i", {\ar"d"},
"m", {\ar"i"},
"l", {\ar"i"},
"i", {\ar"k"},
"i", {\ar"h"},
"f", {\ar"k"},
"c", {\ar"m"},
"m", {\ar"j"},
"h", {\ar"m"},
"l", {\ar"a"},
"h", {\ar"l"},
"k", {\ar"e"},
"k", {\ar"l"},
"a", {\ar"h"},
"h", {\ar"g"},
"j", {\ar"h"},
"g", {\ar"a"},
"b", {\ar"g"},
"g", {\ar"j"},
\end{xy}
\\
(a) && (b) \\
\end{tabular}
\caption{(a) A plabic graph corresponding to the top cell of $\Gr_+(3,7)$ and (b) the associated
dual quiver, associated to a cluster of the $\Gr(3,7)$ cluster algebra.}
\label{fig:one}
\end{figure}
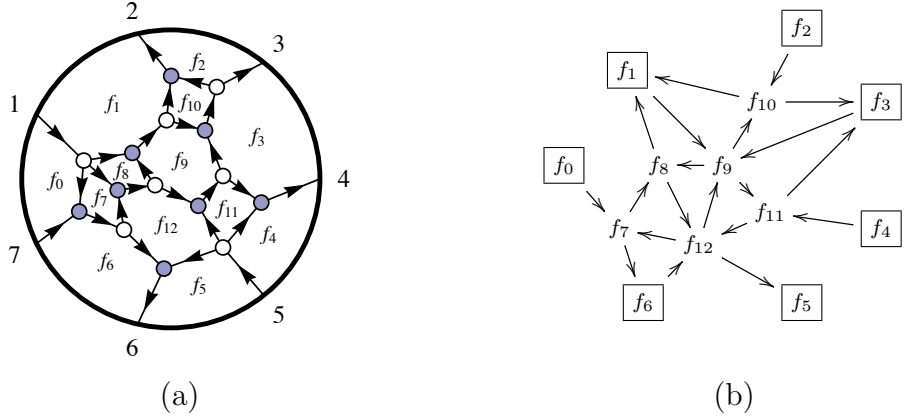

In order to demonstrate clearly what we mean by a ``non-plabic
cluster parameterization'' of a Grassmannian cell, we start with
an example worked out in the style of those considered
in~\cite{Mago:2020kmp}, to which we refer the reader for
all details on notation and conventions.
Whereas
all of the plabic graphs depicted in~\cite{Mago:2020kmp}
had all internal faces bounded by four edges, meaning that
one could always apply a square move to mutate to another plabic graph,
we now consider a plabic graph
which corresponds to the top
cell of $\Gr_+(3,7)$ and
has non-square internal faces, shown in Fig.~\ref{fig:one}(a).
The corresponding boundary measurement is
\begin{align}
\label{eq:cmatrix}
C = \left(\begin{matrix}
1 & c_{12} & c_{13} & c_{14} & 0 & c_{16} & 0 \\
0 & c_{52} & c_{53} & c_{54} & 1 & c_{56} & 0 \\
0 & c_{72} & c_{73} & c_{74} & 0 & c_{76} & 1
\end{matrix}\right)
\end{align}
where
\begin{equation}
\begin{aligned}
c_{12} &= f_0 f_2 f_3 f_4 f_5 f_6 f_{11} f_{12}
(1 + f_7 + f_9 (1 + f_{10}) + f_7 (1 + f_8) f_9 (1+f_{10}))\\
c_{13} &= f_0 f_3 f_4 f_5 f_6 f_{11} f_{12}
(1 + f_7 (1 + f_9 (1+f_8)) + f_9) \\
c_{14} &= f_0 f_4 f_5 f_6 f_{11} f_{12} (1 + f_7) \\
c_{16} &= -f_0 f_6 \\
c_{52} &= f_2 f_3 f_4 f_{11} \\
c_{53} &= f_3 f_4 f_{11} \\
c_{54} &= f_4(1 + f_{11}) \\
c_{56} &= f_0 f_1 f_2 f_3 f_4 f_6 f_7 f_8 f_9 f_{10} f_{11} f_{12} \\
c_{72} &= -f_2 f_3 f_4 f_5 f_6 f_{11} f_{12} (1
+ f_9 (1 + f_{10} )) \\
c_{73} &= -f_3 f_4 f_5 f_6 f_{11} f_{12} (1 + f_9) \\
c_{74} &= -f_4 f_5 f_6 f_{11} f_{12} \\
c_{76} &= f_6
\end{aligned}
\end{equation}
and $\prod f_i = 1$.
For a $7 \times 4$ matrix $Z$,
the solution to $C \, Z = 0$ is
\begin{equation}
\begin{aligned}
f_0&=\frac{\langle 1234\rangle}{\langle 2347\rangle} &
f_1&=\frac{\langle 2367\rangle\langle 2456\rangle}{\langle 6(71)(23)(45)\rangle}\\
f_2&=-\frac{\langle 3456\rangle}{\langle 2456\rangle} &
f_3&=\frac{\langle 1467\rangle\langle 2456\rangle}{\langle 6(71)(23)(45)\rangle}\\
f_4&=-\frac{\langle 1567\rangle}{\langle 1467\rangle} &
f_5&=-\frac{\langle 2346\rangle}{\langle 2345\rangle}\\
f_6&=-\frac{\langle 2347\rangle}{\langle 2346\rangle}&
f_7&=\frac{\langle 1237\rangle\langle 2346\rangle}{\langle 1234\rangle\langle 2367\rangle}\\
f_8&=-\frac{\langle 2347\rangle \langle 6(71)(23)(45)\rangle}{\langle 1237\rangle\langle 2346\rangle\langle 4567\rangle}&
f_9&=\frac{\langle 1267\rangle\langle 2346\rangle\langle 4567\rangle}{\langle 1467\rangle\langle 2367\rangle\langle 2456\rangle}\\
f_{10}&=-\frac{\langle 6(71)(23)(45)\rangle}{\langle 1267\rangle\langle 3456\rangle}&
f_{11}&=-\frac{\langle 6(71)(23)(45)\rangle}{\langle 1567\rangle\langle 2346\rangle}\\
f_{12}&=-\frac{\langle 1467\rangle\langle 2345\rangle\langle 2367\rangle}{\langle 2347\rangle \langle 6(71)(23)(45)\rangle}
\end{aligned}
\label{eq:one}
\end{equation}
where $\langle ijkl \rangle$ are the maximal minors of $Z$ and $\langle a(bc)(de)(fg)\rangle = \langle bade\rangle\langle cafg\rangle - (b \leftrightarrow c)$.

At the level of the plabic graph shown in Fig.~\ref{fig:one}(a), we cannot perform
a move on face $f_{12}$ since it is not a square, but we certainly can mutate on
node $f_{12}$ in the corresponding quiver shown in Fig~\ref{fig:one}(b).
This transforms five of the mutable variables in this cluster according to
\begin{equation}
\begin{aligned}
\begin{split}
f_7 &\mapsto f_7' = f_7/(1 + 1/f_{12}) \,, \\
f_8 &\mapsto f_8' = f_8 (1 + f_{12}) \,,
\end{split}~~
\begin{split}
f_{12} &\mapsto f_{12}' = \frac{1}{f_{12}} \,,
\end{split}~~
\begin{split}
f_9 &\mapsto f_9' = f_9/(1 + 1/f_{12}) \,, \\
f_{11} &\mapsto f_{11}' = f_{11} (1 + f_{12}) \,,
\end{split}
\end{aligned}
\label{eq:two}
\end{equation}
leaving the others unchanged.
If we perform this transformation
on the $C$-matrix shown in~(\ref{eq:cmatrix}), we obtain a new
matrix $C'$ that is not the boundary measurement of any plabic
graph, but perfectly well parameterizes the top cell of
$\Gr_+(3,7)$ as the $f'$s range
over $\mathbb{R}^+$.
Moreover, it is a cluster parameterization in the sense that the
$f'$s are cluster variables of $\Gr(3,7)$ (they belong
to the cluster obtained by
mutating Fig.~\ref{fig:one}(b) on node $f_{12}$).
This exemplifies what we mean by a ``non-plabic
cluster parameterization'', or (more simply) a ``non-plabic
$C$-matrix''.

At the level of the solution to
$C \, Z = 0$, the transformation~(\ref{eq:two}) has the effect of introducing
one additional symbol letter not already present
as a multiplicative factor
in~(\ref{eq:one}). Specifically, by computing
\begin{align}
1 + f_{12} = - \frac{\langle 2346\rangle\langle 7(61)(23)(45)\rangle}{\langle 2347\rangle\langle 6(71)(23)(45)\rangle}
\end{align}
we see that the new factor is $\langle 7(61)(23)(45)\rangle$.

\subsection{Rational Eight-Particle Symbol Letters}

The symbol alphabet for the two-loop NMHV octagon contains~\cite{Zhang:2019vnm}
180 cluster variables of $\Gr(4,8)$:
\begin{itemize}
\item 68 Pl\"ucker coordinates of the form $\an{a\ a{+}1\ b\ c}$,
\item 8 cyclic images of $\an{12\bar{4}\cap\bar{7}}$,
\item 40 cyclic images of $\an{1(23)(45)(78)}$, $\an{1(23)(56)(78)}$, $\an{1(28)(34)(56)}$, $\an{1(28)(34)(67)}$, $\an{1(28)(45)(67)}$,
\item 48 dihedral images of $\an{1(23)(45)(67)},\ \an{1(23)(45)(68)},\ \an{1(28)(34)(57)},$
\item 8 cyclic images of $\an{\bar{2}\cap(245)\cap\bar{8}\cap(856)},$
\item and 8 distinct dihedral images of $\an{\bar{2}\cap(245)\cap\bar{6}\cap(681)}$.
\end{itemize}
We see that 96 are quadratic in Pl\"uckers and
the last
16 are cubic. (The $\Gr(4,8)$ cluster algebra has, in total,
120 quadratic and 174 cubic cluster variables~\cite{JLi}.)
By applying the algorithm described in~\cite{Mago:2020kmp,He:2020uhb}
to all plabic graphs associated to $4k$-dimensional
cells of $\Gr_+(k,8)$, with $1\leq k\leq4$ and including all members of each cyclic class, one encounters all of the Pl\"ucker
coordinates and quadratic cluster variables (in addition, of course,
to numerous non-cluster variables, similar to the examples
described in~\cite{Mago:2020kmp}, as well as the 18 algebraic
symbol letters). However, the cubic symbol letters on the above
list are missing (see also~\cite{HL}).

\begin{figure}
\centering
\includegraphics[width=1.8in]{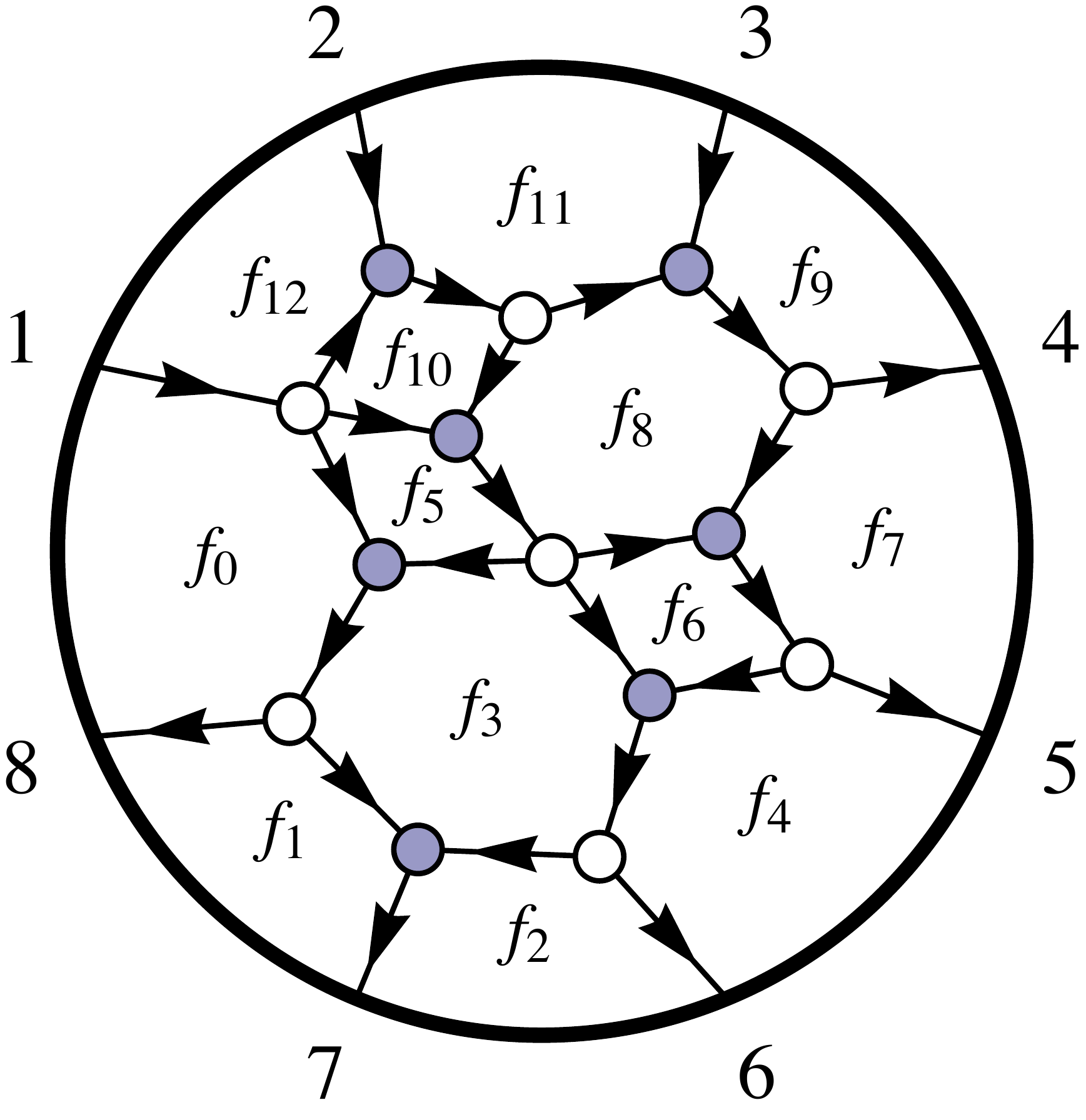}
\caption{A plabic graph associated to the 12-dimensional
cell in $\Gr_+(3,8)$ labeled by the decorated
permutation $\{4,5,7,6,9,8,11,10\}$. The cubic symbol letter
$\an{\bar{4}\cap(467)\cap\bar{2}\cap(278)}$ appears in the solution
of $C \, Z = 0$ after performing a non-square move mutation on face $f_8$.}
\label{fig:two}
\end{figure}

We find that the cubic letters are obtainable if one allows
non-plabic $C$-matrices as described in the previous
subsection. For example, the first type
of cubic letter can be obtained from the (non-top cell) $\Gr_+(3,8)$ plabic
graph shown in Fig.~\ref{fig:two} by applying a mutation on face $f_8$.
We spare the details of writing down the $C$-matrix for this
graph and the solution for all face variables; it suffices to
display
\begin{align}
1+f_8=\frac{\an{3567}\an{\bar{4}\cap(467)\cap\bar{2}\cap(278)}}{\an{2378}\an{4567}\an{3(12)(45)(67)}}
\end{align}
which contains a cubic letter in the same cyclic
class as $\an{\bar{2}\cap(245)\cap\bar{8}\cap(856)}$.

It is also possible to obtain these cubic letters from non-plabic
cluster parameterizations of the top cell of $\Gr_+(4,8)$.
The second type of cubic letter
($\an{\bar{2}\cap(245)\cap\bar{6}\cap(681)}$ and its images) can \emph{only}
be obtained from non-plabic
cluster parameterizations of the top cell.
In the next section we observe that
\emph{all} cluster variables of $\Gr(n{-}m,n)$ can be obtained from
the top cell of $\Gr_+(m,n)$ if one allows consideration
of non-plabic
cluster parameterizations.

\section{A Cluster Isomorphism for Top Cells}
\label{sec:G2n}

In~\cite{Mago:2020kmp,He:2020uhb}
it was noted that in some cases,
the solution to $C \, Z = 0$
(for $C$ a boundary measurement of a plabic graph
corresponding to an $mk$-dimensional
cell of $\Gr_+(k,n)$ and $Z$ an $n \times m$ matrix)
involves \emph{non}-cluster variables of $\Gr(m,n)$.
Moreover, we noted in the previous section
that it is impossible to obtain
all cluster variables of $\Gr(m,n)$ in this way.

Here we point out that if we take $k = n{-}m$, which means
we restrict
our attention to the top cell of $\Gr_+(k,n)$, but allow
arbitrary \emph{non}-plabic $C$-matrices, then
solving $C \, Z = 0$ will always produce
all cluster variables of $\Gr(m,n)$ (which might be
infinite in number), and will never produce non-cluster variables.

\bigskip

To begin with let us consider the simple case $k=2$.
We will demonstrate that the map induced by solving
$C \, Z = 0$ maps Pl\"ucker coordinates of $\Gr(2,n)$ to those
of $\Gr(n{-}2,n)$.
We do not need to choose any particular parameterization, and simply
denote the entries of the $2 \times n$ matrix $C$ by $c_{1,i}, c_{2,i}$ for $i \in [n]$, and let $Z$ be an $n \times (n{-}2)$ matrix with rows $Z_i$. We can choose to solve $C \, Z=0$ for $2n{-}4$ of the $c$ variables in terms of the remaining 4. (Here is the step where it is important that $C$ parameterizes
a top cell; otherwise there would not be enough independent
variables to solve for!). For example, we can express
\begin{align*}
c_{a,i} = (-1)^{n-i-1} \left( c_{a,n{-}1} \frac{\widetilde{\Delta}(i,n)}{\widetilde{\Delta}(n-1,n)} + c_{a,n} \frac{\widetilde{\Delta}(i,n-1)}{\widetilde{\Delta}(n-1,n)}\right) \qquad a \in [2], i \in [n{-}2]
\end{align*}
in terms of the Pl\"ucker coordinates on
$\Gr(n-2,n)$, where $\widetilde{\Delta}(i,j)$
denotes the determinant of $Z$ with rows $i$ and $j$ removed i.e. $\widetilde{\Delta}(i,j)=\epsilon_{i,j,\{1,\dots,n\}\backslash \{i,j\}}\an{\{1\cdots n\}\backslash\{i,j\}}$.
Evaluating the Pl\"ucker coordinates of $\Gr(2,n)$ on the solution
yields
\begin{align}
    \label{eq:G2npluckers}
    \Delta(i,j) = \begin{vmatrix} c_{1i} & c_{1j}\\
                                  c_{2i} & c_{2j}\end{vmatrix} = \widetilde{\Delta}(i,j) \frac{\Delta(n-1,n)}{\widetilde{\Delta}(n-1,n)}\,.
\end{align}
It is worth pointing out that the form of this equation is independent of our choice of solved and unsolved variables, and can be written more invariantly as
\begin{align}
\frac{\Delta(i,j)}{\Delta(k,l)} = \frac{\widetilde{\Delta}(i,j)}{\widetilde{\Delta}(k,l)}
\end{align}
for all $i < j$ and $k< l \in [n]$.
Therefore, up to a single irrelevant overall factor,
we can simply say that $\Delta(i,j) \mapsto \widetilde{\Delta}(i,j)$.
This is the sense in which we say that
solving $C \, Z=0$ maps the Pl\"ucker coordinates of $\Gr(2,n)$ to
those of $\Gr(n{-}2,n)$.

\bigskip

The story is similar for arbitrary $k$.
Let $C$ be a $k \times n$ matrix with entries $c_{ij}$
and let $Z$ be an $(n{-}k) \times n$ matrix.
In order to solve $C \, Z = 0$ let us choose
to solve for
$
    \begin{pmatrix}
    c_{1\,k+1} & \dots & c_{1\,n}\\
    \vdots & \dots & \vdots\\
    c_{k\,k+1} & \dots & c_{k\,n}
    \end{pmatrix}
$ in terms of  $
    \begin{pmatrix}
    c_{1\, 1} & \dots & c_{1\, k}\\
     \vdots & \dots & \vdots\\
    c_{k\, 1} & \dots & c_{k\, k}
    \end{pmatrix}
$. The solution is
\begin{align}
    \label{eq:solgkn}
    c_{a \, b} = (-1)^{b-k} \sum_{i \in U} \frac{\an{i, \lbrace k+1, \dots n \rbrace\backslash b}}{\an{k+1 \dots n}} \, c_{a \,i}
\end{align}
where $U = \lbrace 1, \dots k \rbrace$ and the $\an{\dots}$ represent the Pl\"ucker coordinates of $\Gr(n{-}k,n)$. The Pl\"ucker coordinates of $\Gr(k,n)$, computed from the $C$-matrix, evaluate to
\begin{align}
    \label{eq:plcukersgkn}
\Delta(b_1, \, \dots , \, b_k ) =\epsilon_{b_1,\dots,b_k,\{1,\dots,n\}\backslash \{b_1,\dots,b_k\}} \frac{\an{\lbrace 1, \dots, n\rbrace \backslash \lbrace b_1, \, \dots, \, b_k\rbrace }}{\an{k+1, \, \dots , \, n}} \Delta(1, \, \dots k)\,.
\end{align}
If we choose to label the Pl\"uckers of $\Gr(n{-}k,n)$ by the missing $k$ columns and define
\begin{align}
    \an{a_1 \dots a_{n-k}} \equiv \widetilde{\Delta}\left(\lbrace 1, \dots n\rbrace \backslash \lbrace a_1, \dots a_{n-k} \rbrace\right),
\end{align}
we can rewrite~(\ref{eq:plcukersgkn}) as
\begin{align}
    \frac{\Delta(b_1, \dots b_k)}{\Delta(1, \dots k)} = \frac{\widetilde{\Delta}(b_1, \dots, b_k)}{\widetilde{\Delta}(1, \dots k)}
\end{align}
analogous to~(\ref{eq:G2npluckers}).
Let us recall that the map $\Delta(b_1,\ldots,b_k) \to
\widetilde{\Delta}(b_1,\ldots,b_k)$ induces the `obvious'
isomorphism between the $\Gr(k,n)$ and $\Gr(n-k,n)$ cluster algebras,
in that it maps clusters of one to the other and commutes with mutation.

To conclude:
we have seen that
when $C$ is any cluster parameterization of the top cell of $\Gr_+(n{-}m,n)$,
then the map induced by solving $C \, Z = 0$
is the natural cluster isomorphism $\Gr(n{-}m,n) \mapsto \Gr(m,n)$.
In other words, if $C$ is a top-cell parameterization associated
to some cluster of $\Gr(n{-}m,n)$, 
then the letters appearing in the solution to $C \, Z = 0$ will
be the cluster variables of the image of that cluster in
$\Gr(m,n)$ under the replacement $\Delta \mapsto \widetilde{\Delta}$.
Therefore it is also manifest that we will never
encounter non-cluster variables of $\Gr(m,n)$,
in contrast to what happens
for lower-dimensional cells, such as the examples encountered
in~\cite{Mago:2020kmp,He:2020uhb}.
Fig.~1 of~\cite{Mago:2020kmp} illustrates the
one-to-one correspondence between `input' clusters
of $\Gr(n{-}m,n)$ and `output' clusters
of $\Gr(m,n)$
for the case $m=4$, $n=6$.

\section{Discussion}

The connection between symbol alphabets of scattering amplitudes
in SYM theory and solutions of matrix equations of the form
$C \, Z = 0$ was investigated in~\cite{Mago:2020kmp,He:2020uhb}.
Here we have pointed out that if one allows
$C$ to be an arbitrary cluster parameterization of the
top cell of $\Gr_+(n{-}m,n)$, then \emph{all} cluster variables of
$\Gr(m,n)$ can be obtained from such solutions; moreover,
\emph{no} non-cluster variables arise in this way.
Since all (currently known) rational symbol letters are cluster
variables of $\Gr(4,n)$, we see that they can all be obtained
from the top cell of $\Gr_+(n{-}4,n)$.

One of the main points of~\cite{Mago:2020kmp,He:2020uhb} was that since
the number of plabic graphs is manifestly finite (for any given $n$),
they might select `preferred' finite sets of cluster variables
to serve as candidate symbol alphabets.
This is in line with general expectations~\cite{Prlina:2018ukf,Arkani-Hamed:2019rds} that for every $n$,
there is a finite $n$-particle symbol alphabet that serves to
express all $n$-particle amplitudes to any finite order in perturbation theory.
We (and~\cite{HL}) have found that for $n=8$ (and presumably, also for any higher $n$),
this finite set is \emph{too small}---it is missing 16 of the
rational symbol letters of~\cite{Zhang:2019vnm}.
In this paper we overcame that problem at the expense of introducing
a new one: by allowing arbitrary cluster parameterizations $C$,
we obtain all of the infinitely many cluster
variables of $\Gr(4,n\ge 8)$, not a finite subset. However, we note that while these 16 letters are related to non-plabic cluster parameterizations, the diagrams from which these letters arise in the Landau analysis are still planar~\cite{Dennen:2015bet,Dennen:2016mdk,Prlina:2017azl,Prlina:2017tvx,Prlina:2018ukf}.

It would be interesting to investigate some possibility in the middle---something small enough to give finite sets of cluster variables, but large enough to include all known symbol letters.
We could speculate that our story
bears some superficial resemblance to the tropical fans
and polytopes studied in~\cite{Arkani-Hamed:2019rds,Drummond:2019cxm,Henke:2019hve}.
There, one has the freedom to select from a menu of tropical fans
or dual polytopes of various amounts of fineness, with different
choices being associated to different symbol alphabets.
In the $C \, Z = 0$ story one could imagine restricting attention
to various classes of parameterizations $C$,
in between the extremes of `only plabic graphs' (too few) and `all cluster
parameterizations' (too many).
It would be interesting to investigate this further.

\acknowledgments

We are grateful to N.~Arkani-Hamed for numerous encouraging discussions
and to S.~He for Z.~Li for helpful correspondence.
This work was supported in part by the US Department of Energy under contract
{DE}-{SC}0010010 Task A and by Simons Investigator Award \#376208 (AV).
Plabic graphs were drawn with the help of~\cite{Bourjaily:2012gy}.

\end{document}